\documentclass[12pt]{iopart}
\usepackage{iopams}
\usepackage{graphicx}
\usepackage{setstack}

\begin{document}

\title{Fiber-diffraction Interferometer using Coherent Fiber Optic Taper}

\author{Hagyong Kihm$^*$ and Yun-Woo Lee}
\address{Center for Space Optics, Korea Research Institute of Standards and Science, 1 Doryong-dong, Yuseong-gu, Daejeon 305-340, South Korea}
\ead{hkihm@kriss.re.kr}

%%%%%%%%%%%%%%%%%%% abstract and OCIS codes %%%%%%%%%%%%%%%%

\begin{abstract}

We present a fiber-diffraction interferometer using a coherent fiber optic taper for optical testing in an uncontrolled environment.
We use a coherent fiber optic taper and a single-mode fiber having thermally-expanded core.
Part of the measurement wave coming from a test target is condensed through a fiber optic taper and spatially filtered from a single-mode fiber to be reference wave.
Vibration of the cavity between the target and the interferometer probe is common to both reference and measurement waves, thus the interference fringe is stabilized in an optical way.
Generation of the reference wave is stable even with the target movement.
Focus shift of the input measurement wave is desensitized by a coherent fiber optic taper.
\end{abstract}

\vspace{2pc}
\noindent{\it interferometers}: Article preparation, IOP journals
\submitto{\MST}

\maketitle 

%%%%%%%%%%%%%%%%%%%%%%%%%%  body  %%%%%%%%%%%%%%%%%%%%%%%%%%
\section{Introduction}

We describe a new fiber-diffraction interferometer using a coherent fiber optic taper for stabilizing interference fringes in an uncontrolled environment. Manufacturing technology of optical components such as lenses and mirrors has evolved from the traditional polishing process to the automated machining process.
Diamond turning machine is now able to achieve sub-$\mu$m accuracy over a few hundreds of millimeters in diameter.
And injection molding technology produces high quality optical components with unprecedented yield rate.
Testing those optics depends on optical interferometers due to their fast 2 dimensional measurement and quality inspection at the operational wavelength of the test optics.
But the interferometer is so vulnerable to the external vibration that isolating the test table from the machining bed is essential in most cases.
This fact deters the interferometer system from being united with the machining center as a truly repeatable feedback sensor.
Nowadays, researches on vibration insensitive (anti-vibration) interferometers have been spurred to meet those industrial demands, including large scale optics used for ignition facility and telescope optics~\cite{Wyant2003}.
Several techniques are used for desensitizing the interferometer to the external vibration and can be categorized into 3 classes; common-path configuration, vibration feedback control, and spatial phase shifting for real time inspection.

Common-path enables the vibration of test optics common to both reference and measurement waves, whereby interference fringe between them looks static.
Point diffraction interferometer by Smartt and Steel~\cite{Smartt1975} is an example, which makes reference wave from the focused measurement wave through a pinhole.
Lateral shearing interferometer is a common-path interferometer and optical pick-up lenses in production line were tested by Cho and Kim~\cite{Cho1997}. 
Scatter plate interferometer is also a common-path interferometer with high immunity to external vibration~\cite{Morris2002}. 

Another technique for anti-vibration is direct feedback control.
Yoshino and Yamaguchi~\cite{Yoshino1998} implemented a closed-loop phase shifting Fizeau interferometer where optical phases are detected by a two-frequency optical heterodyne method.
Twymann-Green interferometer with a single point detector and an electro-optic modulator was also proposed~\cite{Yamaguchi1996}.

Last category of anti-vibration technique is a single-shot interferometer with spatial phase shifting.
Phase shifting is necessary to enhance measurement accuracy and this is accomplished spatially as opposed to temporal phase shifting techniques.
Smythe and Moore~\cite{Smythe1984} used polarization beam splitters and waveplates to acquire 4 phase-shifted fringes. 
Millerd et al.~\cite{Millerd2001} used holographic elements and polarizers to obtain 4 phase-shifted fringe. 
Pixelated phase-mask can also be used for spatial phase shifting~\cite{Millerd2004a}. 
Spatial carrier phase shifting technique is widely used due to its simple and easy embodiment~\cite{Melozzi1995}. 
By introducing linear tilt phase term in reference or test beam, a spatially modulated fringe can be obtained and analyzed with several algorithms such as sinusoidal fitting~\cite{Ransom1986} and Fourier analysis~\cite{Takeda1982}. 

Aforementioned anti-vibration interferometers of common-path, closed-loop feedback control, and spatial phase shifting are vying each other with their relative advantages and disadvantages.
Closed-loop feedback cannot control high frequency vibrations over the bandwidth of control loop and the actual implementation is difficult.
Spatial phase shifting suffers the same problem with high frequency vibrations because the detector frame rate and shutter speed is generally limited. 
And highly repeatable measurement is practically impossible between successive tests, which lowers the system reliability.
Common-path interferometers are ideal for stabilizing fringes in principle, but they usually lack real time capability due to inherent temporal phase shifting principles.

Combining common-path configuration with a spatial phase shifting technique can be a promising solution to anti-vibration interferometry.
For examples, Kwon~\cite{Kwon1984} made three phase-shifted fringes with a phase grating and a pinhole achieving a common-path real time interferometer. 
Millerd et al.~\cite{Millerd2004b} from 4D Technology Corporation combined a point diffraction interferometer with their spatial phase shifting interferometer. 
Both interferometers use pinholes and their diffracting fields as reference waves.
Recently, Kihm and et al.~\cite{Kihm2005} reported this type of interferometer using a single-mode fiber, but it suffers from difficulties of focusing aberrant measurement beam into the small core of a single mode fiber.
Defocus or lateral vibration of the target severely affects beam intensity out of the single-mode fiber. 
Therefore fringe visibility changes when large vibration is involved.
This paper overcomes that weakness and introduces a new type of interferometer.

The main idea of the research is condensing measurement wave through a coherent fiber optic taper\,(FOT) and a single-mode fiber\,(SMF) with thermally-expanded core\,(TEC) to make reference wave.
Vibration of the cavity between the target and the interferometer probe is common to both reference and measurement waves, thus the interference fringe is stabilized in an optical way.
Generation of the reference wave is stable even with the target movement.
Focus shift of the input measurement wave is desensitized by an FOT.
The uncertainly of measurement results can be lowered due to highly repeatable performance even with external vibrations.
Principles will be explained in Sec.~\ref{sec:princ} and experimental results will be detailed in Sec.~\ref{sec:expr} followed by conclusions in Sec.~\ref{sec:conc}.

\section{Principles}
\label{sec:princ}
\begin{figure}[tbp]
  \centering
  \includegraphics[width=2.4in]{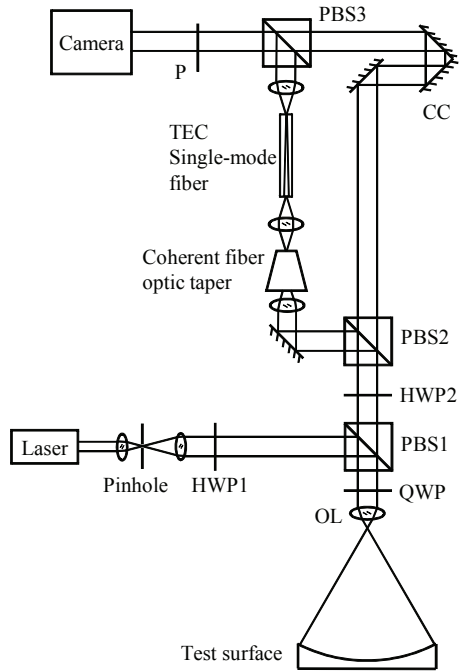}
  \caption{Principles of fiber-diffraction interferometer using an FOT: HWP, half-wave plate; QWP, quarter-wave plate; PBS, polarization beam splitter; OL, objective lens; CC, corner cube; P, polarizer.}
  \label{fig:fig01}
\end{figure}

The fiber-diffraction interferometer using an FOT is shown in Fig.~\ref{fig:fig01}.
Any type of laser, which is linearly polarized and spatially coherent, can be used as a light source.
Continuous wave lasers like He-Ne lasers or super luminescent diodes (SLD) of short coherence length can be used for general purposes. Pulse lasers could be used for stroboscopic inspection.
The laser is filtered through a pinhole and collimated by a lens becoming well-defined plane wave.
Half-wave plate 1 (HWP1) rotates the polarization angle of the beam and controls the amount of reflected beam at the polarization beam splitter 1 (PBS1).
This eventually adjusts the brightness of interference fringes.
A quarter-wave plate (QWP) with its fast axis aligned at $45^{\circ}\/$ makes the beam circularly-polarized.
Objective lens (OL) or null lens in case of aspheric targets forms the beam wavefront fit to the target surface for reciprocation.
The reflected measurement wave, which is circularly polarized, is then linearly polarized in orthogonal direction after the QWP and passes through the PBS1.
HWP2 rotates the polarization angle of the measurement wave, thus controls the split ratio between two arms at PBS2.
The reference arm is composed of a focusing lens, an FOT and an SMF with TEC to make spatially filtered wavefront.
Corner cube (CC) in the measurement arm translates to compensate optical path length difference between two arms.
PBS3 combines those two beams and polarizer (P) with $45^{\circ}\/$ filters them in diagonal direction making interference fringes at the detector.

When phase shifting is required for wavefront analysis, detectors equipped with a spatial phase shifter~\cite{Millerd2001,Millerd2004a,Kihm2005} could be used. Translation of CC for temporal phase shifting might be adopted in vibration-isolated environment.
In this paper, we focus mainly on the verification of using an FOT in fiber-diffraction interferometry.
Also we assume that optical parts comprising the interferometer are fixed as a single body.
Relative motion or vibration of the test surface doesn't affect the alignment of the interferometer components.
This can be justified in general optical testing environment where unstable cavity between the probe and a target is the major contributor lowering repeatability.

Fringe stabilization is possible even with external vibrations by making reference wave out of measurement wave.
Shared vibrant phase which is mainly piston motion cancel out and the fringes look static.
A pinhole has been used for this purpose in point diffraction interferometry~\cite{Smartt1975,Medecki1996}.
They use a sidelobe of focused measurement wave and tilted fringes are inevitable.
Demodulating fringes~\cite{Ransom1986,Takeda1982} lowers measurement accuracy and high frequency features cannot be recovered.
An SMF can be used due to the high quality wavefront output and easy embodiment~\cite{Kihm2005}.
But the difficulty of focusing aberrant measurement wave into the small core of a fiber poses doubt about practical uses.
Increasing coupling efficiency of a laser source into an SMF has been a major research activity in optical communication, and thermally expanding the core(TEC) of a fiber is an example~\cite{Hanafusa1991}.

In this research we propose to use an FOT as a beam coupler for interferometric uses.
An FOT consists of a large number of optical fibers fused together to form a coherent bundle.
The bundle is heat formed, resulting in variation of its diameter from one end to the other.
The magnification of a taper is simply the ratio of the diameter of the end faces, which is generally 2\textendash5$\times$~\cite{schott}. 
The light transmission of an FOT is given in terms of internal transmission of glass core, Fresnel reflection at the end faces, and the ratio of core area to the total area termed packaging fraction (PF).
The transmission $T$ can be expressed as
\begin{equation}
  \label{eq:T}
  T = \mathrm{PF}t_{f}\exp\left(-\beta_{\lambda}L\right)
\end{equation}
, where $t_{f}$ is the Fresnel transmission factor, $\beta_{\lambda}\/$ is the absorption coefficient of the core glass, and $L$ is the length of the taper~\cite{Peli1997}.
The light transmission can be increased by anti-reflection (AR) coating on input and output ports of the taper.
A pinhole or a coated mask on the small end face blocks unwanted spurious modes~\cite{Shi2006} and passes only a single-mode by point diffraction.

Immunity to the focus shift at the input port is achieved by condensing light through an FOT.
The input end face of a taper is placed near the aperture stop of a focusing objective lens.
This slightly defocused input beam looks static at the output even when the vibration of the target changes the direction\,(tilt) and divergence angle\,(focus) of the measurement wave.
High numerical aperture\,(NA) of the FOT, which is 1, captures almost every incoming field from the objective lens.
The beam spot size at the output port is reduced according to the reduction ratio of the taper.
Multi-mode fields at the output, however, should be filtered into a single-mode to be used as reference wave~\cite{Li1985,Leon-Saval2005}.
A TEC-processed SMF is used to combine individual fibers at the output of the FOT.
Then, the reference wave from the SMF is quite stable in its amplitude while carrying vibrant phase motion of the target.

Considering the power of available lasers and sensitivity of detectors, transmission loss through an FOT and a TEC-processed SMF is not a problem in interferometric applications~\cite{Kosterin2004}.  
The following section explains the actual implementation and verifies the use of an FOT in fiber-diffraction interferometry.

\section{Experimental Results and Discussion}
\label{sec:expr}
Reference wave is made by fiber diffraction from the end face of an SMF.
Focusing aberrant measurement wave into the small core of SMF is difficult when the test optic has vibrational motions.
Axial motion makes the beam defocused at the input port.
Lateral or tilt motion shifts the beam focus and lowers the coupling efficiency more severely.
When large vibration is involved at the cavity between the test optic and the interferometer probe, the amplitude of reference wave will fluctuate accordingly.
Thus we cannot get stable fringes enough for practical phase measurement.
The objective of this research is to get stable fringes immune to the focus shift of the measurement wave.

\begin{figure}[tbp]
  \centering
  \includegraphics[width=2.4in]{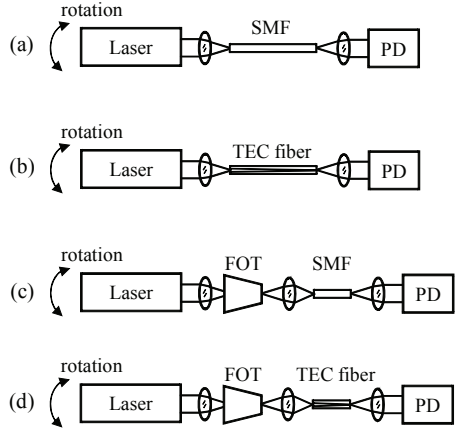}
  \caption{Coupling efficiency test with respect to input focus shift: (a) single-mode fiber\,(SMF) only, (b) thermally-expanded core\,(TEC) fiber only, (c) fiber-optic taper\,(FOT) and SMF, (d) FOT and TEC fiber; PD, photo-detector.}
  \label{fig:fig02}
\end{figure}

We verified the improvement of using an FOT by comparative experiments shown in Fig.~\ref{fig:fig02}.
The laser is focused into an SMF and a photo-detector\,(PD) captures the output intensity to examine the coupling efficiency.
The laser is rotated to simulate tilt motions of the measurement wave.
We tested with 4 different set-ups; (a) using an SMF, (b) a TEC fiber, (c) an FOT and an SMF, and (d) an FOT and a TEC fiber.
Fig.~\ref{fig:fig02}(a) is a conventional method using an SMF in fiber-diffraction interferometry.
TEC fiber in Fig.~\ref{fig:fig02}(b) has a longitudinal variation of the core diameter.
Mode-field diameter of the input end becomes 10\,$\mu$m from 4\,$\mu$m after the TEC process.
This makes the coupling less sensitive to the beam focus.
We used an FOT from Schott~\cite{schott}.
The diameter of a large end is 25\,mm  and the small end is 8\,mm. The magnification is 3.1:1, which is the ratio of end face diameters.
The element size of each fiber at large end side is 6\,$\mu$m and the ratio of core and cladding area is 1:1.
The PF in Eq.~\ref{eq:T} is lower than 50\% due to the extramural absorption (EMA) which eliminates stray light through the cladding~\cite{Siegmund1966}.
The refractive index of the core is 1.810 and NA of a fiber at the end face is 1 from manufacturer's specification~\cite{schott}.

\begin{figure}[tbp]
  \centering
  \includegraphics[width=3in]{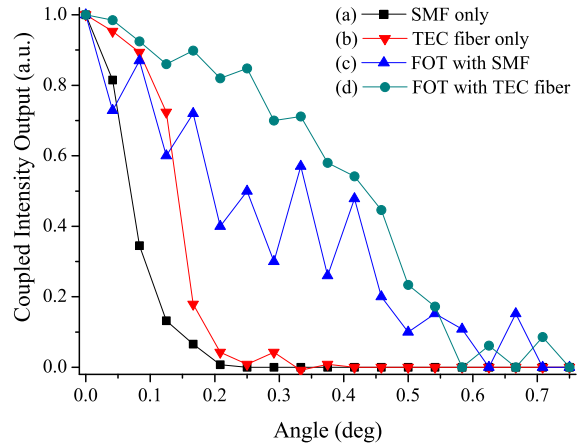}
  \caption{Coupled intensity outputs from the experiments shown in Fig.~\ref{fig:fig02}}
  \label{fig:fig03}
\end{figure}

The output is a convolution process between the focused laser and the input modal field of fibers and the FOT.
The NA of the SMF and the TEC fiber is 0.12, while the NA of the FOT is 1.
The mode-field diameters of the SMF and the TEC fiber are 4\,$\mu$m and 10\,$\mu$m respectively.
In case of (c) and (d) in Fig.~\ref{fig:fig02}, mode-field diameter depends on the magnifications of the lens and the FOT.
We used a lens with 1$\times$ magnification and an FOT with 3.1$\times$ magnification.
The input mode-field diameters of Fig.~\ref{fig:fig02}(c) and Fig.~\ref{fig:fig02}(d) are 12\,$\mu$m and 31\,$\mu$m respectively.

Fig.~\ref{fig:fig03} shows the experimental results with the set-up shown in Fig.~\ref{fig:fig02}.
Coupled intensity output from the PD is normalized in each case.
Immunity to the focus shift can be compared by their relative output profiles, not their maximum values.
If the plot is steep in a small input variation, the sensitivity to the focus shift is high.
The SMF is the most sensitive as shown in Fig.~\ref{fig:fig03}(a).
The TEC fiber in Fig.~\ref{fig:fig03}(b) is relatively insensitive to the focus shift due to the enlarged mode-field diameter.
The FOT has a large NA accepting almost every incoming field, but the EMA makes the beam fluctuate as shown in Fig.~\ref{fig:fig03}(c).
This problem can be minimized by using a TEC fiber as in Fig.~\ref{fig:fig03}(d).
The FOT's large NA makes the output robust to the focus shift and the TEC fiber's large mode-field diameter mitigates EMA effect.
About 5 times improvement is observed in terms of acceptance angle when we compare $1/e^2$ intensity outputs of Fig.~\ref{fig:fig03}(a) and Fig.~\ref{fig:fig03}(d).

\begin{figure}[tbp]
  \centering
  \includegraphics[width=4in]{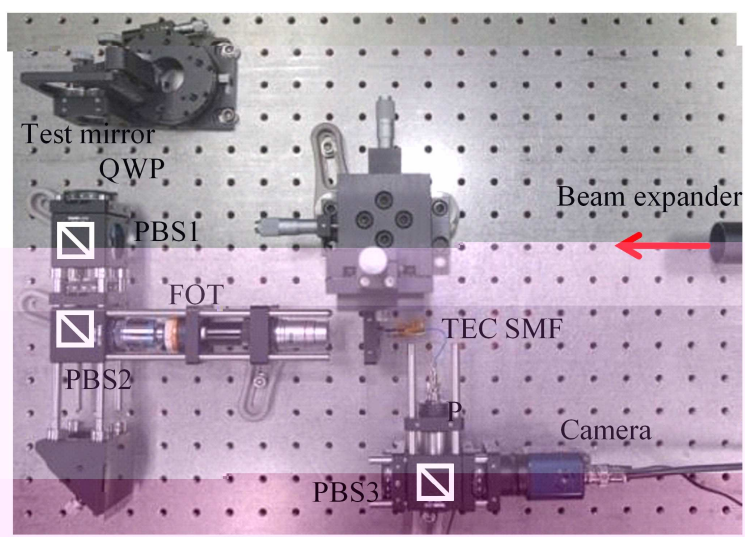}
  \caption{Experimental set-up of the fiber-diffraction interferometer using an FOT in Fig.~\ref{fig:fig01}}
  \label{fig:fig04}
\end{figure}

\begin{figure}[tbp]
  \centering
  \includegraphics[width=4in]{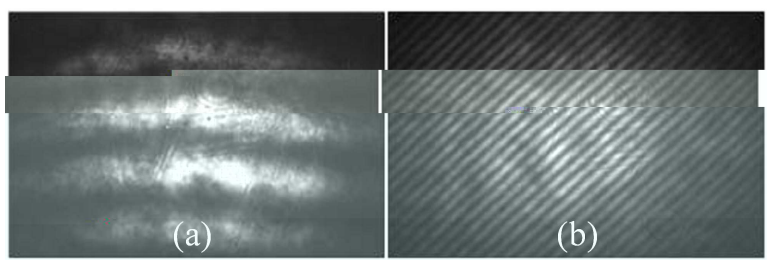}
  \caption{Captured fringes from the set-up in Fig.~\ref{fig:fig04} when the test surface is rotated by the angle (a) $0.02\,^{\circ}$ and (b) $0.2\,^{\circ}$}
  \label{fig:fig05}
\end{figure}

Fig.~\ref{fig:fig04} shows an experimental set-up of the fiber-diffraction interferometer using an FOT and a TEC SMF.
Optical components are arranged similar to the layout of Fig.~\ref{fig:fig01} except for the test surface.
We used a flat mirror instead of a spherical mirror.
The flat mirror is suitable for evaluating immunity to the focus shift due to tilt motions.
The TEC SMF is bent $90\,^{\circ}$ to filter out unwanted multi-modes effectively within a short optical path.
Fig.~\ref{fig:fig05}(a) shows an interferogram when the test surface has a tilt angle of $0.02\,^{\circ}$.
Fig.~\ref{fig:fig05}(b) was obtained when the test surface was rotated by $0.2\,^{\circ}$.
Fringe visibility is comparable to that of Fig.~\ref{fig:fig05}(a).
This is expected in Fig.~\ref{fig:fig03}(d) where the coupled intensity output is over 80\,\% of its maximum value at the angle $0.2\,^{\circ}$.
The fiber-diffraction interferometer using an SMF~\cite{Kihm2005} cannot obtain fringes like Fig.~\ref{fig:fig05}(b).
As evident in Fig.~\ref{fig:fig03}(a), the coupled intensity output from an SMF is almost zero at the angle $0.2\,^{\circ}$.
This result proves the performance improvement with the use of an FOT in fiber-diffraction interferometry.

\section{Conclusions}
\label{sec:conc}

We proposed a new fiber-diffraction interferometer using a coherent fiber optic taper for stabilizing fringes in vibrational optical testing environment.
Reference wave is made by condensing measurement wave through a coherent fiber optic taper and a single-mode fiber with thermally expanded core.
Experimental comparison proved the superior coupling efficiency of the proposed method.
Combining this technique with a spatial phase shifter will increase measurement repeatability as well as freezing vibrations, whereby ideal vibration insensitive interferometer could be realized.
Cascading multiple fiber optic tapers to study coupling efficiency and comparing with the state of the art commercial interferometers will be the future works.

\section{References}


\begin{thebibliography}{10}

\bibitem{Wyant2003}
J.~C. Wyant.
\newblock Dynamic interferometry.
\newblock {\em Optics and Photonics News}, 14:36--41, 2003.

\bibitem{Smartt1975}
R.~N. Smartt and W.~H. Steel.
\newblock Theory and application of point-diffraction interferometers.
\newblock {\em Jap. J. Appl. Phys.}, 14 (Supplement 14-1):351--356, 1975.

\bibitem{Cho1997}
W.-J. Cho and S.-W. Kim.
\newblock Stable lateral-shearing interferometer for production-line inspection
  of lenses.
\newblock {\em Opt. Eng.}, 36:896--900, 1997.

\bibitem{Morris2002}
M.~B. North-Morris, J.~VanDelden, and J.~C. Wyant.
\newblock Phase-shifting birefringent scatterplate interferometer.
\newblock {\em Appl. Opt.}, 41(4):668--677, 2002.

\bibitem{Yoshino1998}
T. Yoshino and H. Yamaguchi.
\newblock Closed-loop phase-shifting interferometry with a laser diode.
\newblock {\em Opt. Lett.}, 23(20):1576--1578, 1998.

\bibitem{Yamaguchi1996}
I. Yamaguchi, J.-Y. Liu, and J.-I. Kato.
\newblock Active phase-shifting interferometers for shape and deformation
  measurements.
\newblock {\em Opt. Eng.}, 35(10):2930--2937, 1996.

\bibitem{Smythe1984}
R.~Smythe and R.~Moore.
\newblock Instantaneous phase measuring interferometry.
\newblock {\em Opt. Eng.}, 23:361--364, 1984.

\bibitem{Millerd2001}
J.~E. Millerd and N.~J. Brock.
\newblock Methods and apparatus for splitting, imaging and measuring wavefronts
  in interferometry.
\newblock U.S. Patent 6,304,330, October 16, 2001.

\bibitem{Millerd2004a}
J.~E. Millerd, N.~J. Brock, J.~B. Hayes, M.~B. North-Morris, M.
  Novak, and J.~C. Wyant.
\newblock Pixelated phase-mask dynamic interferometer.
\newblock {\em Proc. SPIE}, 5531:304--314, 2004.

\bibitem{Melozzi1995}
M. Melozzi, L. Pezzati, and A. Mazzoni.
\newblock Vibration-insensitive interferometer for on-line measurements.
\newblock {\em Appl. Opt.}, 34(25):5595--5601, 1995.

\bibitem{Ransom1986}
P.~L. Ransom and J.~V. Kokal.
\newblock Interferogram analysis by a modified sinusoid fitting technique.
\newblock {\em Appl. Opt.}, 25(22):4199--4204, 1986.

\bibitem{Takeda1982}
M. Takeda, H. Ina, and S. Kobayashi.
\newblock Fourier-transform method of fringe-pattern analysis for
  computer-based topography and interferometry.
\newblock {\em J. Opt. Soc. Am.}, 72(1):156--160, 1982.

\bibitem{Kwon1984}
O.~Y. Kwon.
\newblock Multichannel phase-shifted interferometer.
\newblock {\em Opt. Lett.}, 9(2):59--61, 1984.

\bibitem{Millerd2004b}
J.~E. Millerd, N.~J. Brock, J.~B. Hayes, and J.~C. Wyant.
\newblock Instantaneous phase-shift point-diffraction interferometer.
\newblock {\em Proc. SPIE}, 5531:264--272, 2004.

\bibitem{Kihm2005}
H. Kihm and S.-W. Kim.
\newblock Fiber-diffraction interferometer for vibration desensitization.
\newblock {\em Opt. Lett.}, 30(16):2059--2061, 2005.

\bibitem{Medecki1996}
H.~Medecki, E.~Tejnil, K.~A. Goldberg, and J.~Bokor.
\newblock Phase-shifting point diffraction interferometer.
\newblock {\em Opt. Lett.}, 21(19):1526--1528, 1996.

\bibitem{Hanafusa1991}
H.~Hanafusa, M.~Horiguchi, and J.~Noda.
\newblock Thermally-diffused expanded core fibers for low-loss and inexpensive
  photonic components.
\newblock {\em Electron. Lett.}, 27:1968--1969, 1991.

\bibitem{schott}
SCHOTT North America, Inc. 555 Taxter Road Elmsford, NY 10523 USA. http://www.us.schott.com/lightingimaging/english/. Accessed on Jul. 2010.

\bibitem{Peli1997}
E. Peli and W.~P. Siegmund.
\newblock Fiber-optic reading magnifiers for the visually impaired.
\newblock {\em J. Opt. Soc. Am. A}, 12(10):2274--2285, 1995.

\bibitem{Shi2006}
K. Shi, F.~G. Omenetto, and Z. Liu.
\newblock Supercontinuum generation in an imaging fiber taper.
\newblock {\em Opt. Express}, 14(25):12359--12364, 2006.

\bibitem{Li1985}
Y.-F. Li and J. W.~Y. Lit.
\newblock Transmission properties of a multimode optical-fiber taper.
\newblock {\em J. Opt. Soc. Am. A}, 2(3):462--468, 1985.

\bibitem{Leon-Saval2005}
S.~G. Leon-Saval, T.~A. Birks, J.~Bland-Hawthorn, and M.~Englund.
\newblock Multimode fiber devices with single-mode performance.
\newblock {\em Opt. Lett.}, 30(19):2545--2547, 2005.

\bibitem{Kosterin2004}
A. Kosterin, V. Temyanko, M. Fallahi, and M. Mansuripur.
\newblock Tapered fiber bundles for combining high-power diode lasers.
\newblock {\em Appl. Opt.}, 43(19):3893--3900, 2004.

\bibitem{Siegmund1966}
W.~P. Siegmund.
\newblock Fiber optical image transfer device having a multiplicity of light
  absorbing elements.
\newblock U.S. Patent 3,247,756, April 26, 1966.

\end{thebibliography}
\end{document}